\documentclass{jpp}
\pdfoutput = 1 
\usepackage {graphicx}
\usepackage{graphicx}
\usepackage{epstopdf, epsfig}
\usepackage{hyperref}
\usepackage{amsmath,amsfonts,amssymb}
\usepackage{subfigure}

\begin{document}
\title{Equilibrium of a Rapidly Rotating Axisymmetric Magnetic Mirror Machine}
\author{Richard Fitzpatrick\footnote{rfitzp@utexas.edu}}
\affiliation{Institute for Fusion Studies,  Department of Physics, University of Texas at Austin,  Austin TX 78712, USA}
\maketitle

\begin{abstract}
A recent paper [Hazeltine, et al., Phys.\ Plasmas {\bf 33}, 072501 (2026)] has questioned whether the standard result, (ultimately) due to Ferraro, 
that the plasma angular
velocity is approximately constant along individual equilibrium magnetic field-lines in a rotating axisymmetric magnetic mirror machine, 
continues to hold when the rotation becomes sonic or supersonic. In order to resolve this issue, the equilibrium of a rapidly rotating mirror is investigated,
starting from first principles, using an ideal two-fluid model with anisotropic pressure. It is found that, as long as the ion gyro-radius is much less than the machine size,
and the angular velocity of the plasma is much less than the ion gyro-frequency, the Ferraro result holds good. 
\end{abstract}

\section{Introduction}
The centrifugal mirror conﬁnement concept incorporates sonic, or even supersonic,  plasma rotation  into a magnetic mirror device. This concept has been shown experimentally to significantly decrease parallel losses, and to increase plasma stability, as compared with non-rotating axisymmetric mirrors [\cite{,Aydemir2004,Endrizii2023,Schwatz2024}].

A recent paper [\cite{Hazeltine2026}]  has
questioned whether the standard result, (ultimately) due to Ferraro [\cite{Ferraro1937}], 
that the plasma angular
velocity is approximately constant along individual equilibrium magnetic field-lines in a rotating axisymmetric magnetic mirror machine, 
continues to hold when the rotation becomes sonic or supersonic. 
 In order to resolve this issue, in this paper we  shall investigate the equilibrium of a rapidly rotating mirror, 
starting from first principles, using an ideal two-fluid model.

\section{Anisotropic pressure tensor}
We wish to generalize the analysis of \cite{Hazeltine2026} to include the effects of particle trapping, because this is the
main confinement mechanism both  in a weakly rotating mirror device [\cite{Fitzpatrick2023}] and very close to the axis of  a strongly rotating device
[\cite{Schwatz2024}]. Within the context of fluid theory, we can achieve our goal by adopting anisotropic
pressure tensors for the electrons and the ions [\cite{Chew1956,Togo2018,Zhao2021}]. Let 
\begin{equation}
\underline{\underline{P}}_{\,s} = p_{\parallel\,s}\,{\bf b}\,{\bf b} + p_{\perp\,s}\left({\bf 1} - {\bf b}\,{\bf b}\right)
\end{equation}
be the gyrotropic pressure tensor for species $s$ (the two species being electrons, $s= e$, and ions, $s= i$). 
Here, ${\bf b} = {\bf B}/B$, where ${\bf B}$ is the magnetic field-strength, ${\bf 1}$ is the identity tensor, and $p_{\parallel\,s}$ and $p_{\perp\,s}$
are the species-$s$ parallel and perpendicular (relative to the magnetic field) pressures. In other words,
\begin{align}
p_{\parallel\,s}({\bf r}) &= m_s\int (v_{\parallel\,s}-V_{\parallel\,s})^2\,f_s({\bf r},{\bf v}_s)\,d^3{\bf v}_s,\\[0.5ex]
p_{\perp\,s}({\bf r}) &= \frac{m_s}{2}\int (v_{\perp\,s}-V_{\perp\,s})^2\,f_s({\bf r},{\bf v}_s)\,d^3{\bf v}_s,
\end{align}
where $m_s$ is the species-$s$ mass, $f_s({\bf r},{\bf v}_s)$ the (ensemble averaged) species-$s$ distribution function, ${\bf V}_s$
the species-$s$ fluid velocity, $v_{\parallel\,s}= {\bf b}\cdot{\bf v}_s$, ${\bf v}_{\perp\,s} = {\bf v}_s - v_{\parallel\,s}\,{\bf b}$,
et cetera [\cite{Fitzpatrick2023}].

It is easily demonstrated that [\cite{Schwatz2024}]
\begin{equation}
\nabla\cdot \underline{\underline{P}}_{\,s} = \nabla p_{\parallel\,s}+ (p_{\perp\,s}-p_{\parallel\,s})\nabla_\parallel \ln B
+ \nabla_\perp(p_{\perp\,s} - p_{\parallel\,s}) - (p_{\perp\,s}-p_{\parallel\,s})\,\boldsymbol{\kappa}_\perp,
\end{equation}
where $\boldsymbol{\kappa}_\perp = ({\bf b}\cdot\nabla) {\bf b}$ is the magnetic field-line curvature vector [\cite{Freidberg2014}], $\nabla_\parallel\equiv {\bf b}\,{\bf b}\cdot\nabla$, and
$\nabla_\perp\equiv \nabla-\nabla_\parallel$.  Note that ${\bf b}\cdot\boldsymbol{\kappa}_\perp=0$. 

It is helpful to introduce the concepts of species-$s$ parallel and perpendicular temperatures, via the definitions  $p_{\parallel\,s} = n_s\,T_{\parallel\,s}$ and $p_{\perp\,s} = n_s\,T_{\perp\,s}$, where $n_s$ is the species-$s$ number density. In a completely collisionless, strongly magnetized, plasma with negligible
heat fluxes, we expect $T_{\perp\,s}\propto B$ and $T_{\parallel\,s}\propto n_s^{\,2}/B^{2}$ on a particular magnetic field-line [\cite{Chew1956,Kulsrud1983}].
The first scaling follows directly from the conservation of the magnetic moments of charged particles [\cite{Northrop1963,Fitzpatrick2023}], and is fairly robust, whereas the second depends on the  evolution of the parallel dynamics, and is much easier to violate because it is affected by processes that act along magnetic field-lines.
Generally speaking, we expect $\langle T_{\perp\,s}\rangle > \langle T_{\parallel\,s}\rangle$ on a given magnetic field-line in a mirror device,
where $\langle\cdots\rangle$ denotes an average along the field-line [\cite{Post1987,Ryutov1990}]. One obvious cause of this asymmetry is the conservation of
the magnetic moment, which converts parallel kinetic energy into perpendicular kinetic energy as a charged particle moves along a magnetic field-line
into a region of greater magnetic field-strength. Thus, if we imagine that $T_{\perp\,s}= T_{\parallel\,s}$ on the mid-plane of the device,
where the magnetic field-strength is smallest, then the conservation of magnetic moment ensures that $\langle T_{\perp\,s}\rangle > \langle T_{\parallel\,s}\rangle$ on a given magnetic field-line. Another cause of the asymmetry is the fact that the loss-cone in velocity space is almost empty (in a near-collisionless device), which ensures that 
$T_{\perp\,s}> T_{\parallel\,s}$  even on the mid-plane. Generally speaking, we expect the temperature anisotropy to be less marked
for   electrons than for ions, because the collisional relaxation of the anisotropy takes place much more rapidly for the former species [\cite{Fitzpatrick2023}]. In practice, the ion temperature asymmetry is often also driven by fueling via neutral beam injection directed perpendicular to magnetic field-lines, which preferentially injects ions with
larger perpendicular than parallel kinetic energies into the plasma [\cite{Endrizii2023}]. There is no analogous external driving mechanism for the electron temperature
anisotropy. 

In this paper, we shall adopt a somewhat simplistic model in which $T_{\parallel\,s}$ and $T_{\perp\,s}$ represent the {\em average}\/
values of the species-$s$ parallel and perpendicular temperatures, respectively, on a given magnetic field-line. Thus, by definition,  
$T_{\parallel\,s}$ and $T_{\perp\,s}$ are constant on a given field-line. However, in accordance with the previous discussion, we expect $T_{\perp\,s} >T_{\parallel\,s}$. Moreover,
we shall allow $T_{\parallel\,s}$ and $T_{\perp\,s}$ to vary from field-line to field-line. 
 It follows that
\begin{align}\label{e5x}
\nabla\cdot \underline{\underline{P}}_{\,s} &= n_s\,T_{\parallel\,s}\left[\nabla_\parallel\ln n_s+ \gamma_s\,\nabla_\parallel\ln B +(1+\gamma_s)\,\nabla_\perp\ln (n_s\,T_{\parallel\,s})
 +\nabla_\perp\gamma_s -\gamma_s\,\boldsymbol{\kappa}_\perp\right],
\end{align}
where
\begin{align}
\gamma_s &= \frac{T_{\perp\,s} - T_{\parallel\,s}}{T_{\parallel\,s}}
\end{align}
is a positive constant on a given magnetic field-line. Note that the only two force densities that act parallel to magnetic field-lines on the right-hand side of Eq.~(\ref{e5x}) (i.e., the first two terms within the square parenthesis) are, respectively, associated
with the parallel density gradient and the fluid average of the single-particle mirror force, $-\mu_s\,\nabla B$, where $\mu_s$ is the species-$s$ magnetic moment. Thus, if $\gamma_s$ is positive then the mirror force can maintain a parallel density gradient on a given magnetic field-line in which the density is reduced in regions of
strong magnetic field-strength. Of course, this is the whole basis of conventional mirror confinement. 

\section{Fundamental equations}
Within the context of ideal two-fluid theory, the complete set of equations  that governs the equilibrium of a  mirror device are [\cite{Fitzpatrick2023}]
\begin{align}
\nabla\cdot{\bf E} &= \frac{e\,(n_i-n_e)}{\epsilon_0},\label{e1}\\[0.5ex]
\nabla\cdot{\bf B} &= 0,\\[0.5ex]
\nabla\times{\bf E} &= {\bf 0},\\[0.5ex]
\nabla\times{\bf B} &= \mu_0\,e\,(n_i\,{\bf V}_i-n_e\,{\bf V}_e),\label{e4}
\end{align}
and 
\begin{align}
{\bf 0 } &={\bf V}_e \cdot \nabla n_e+ n_e\,\nabla\cdot{\bf V}_e,\label{e4a}\\[0.5ex]
{\bf 0}&= e\,n_e\,({\bf E}+{\bf V}_e\times{\bf B})+\nabla\cdot \underline{\underline{P}}_{\,e} ,\label{e5}
\end{align}
and
\begin{align}\label{e6a}
{\bf 0 } &={\bf V}_i \cdot \nabla n_i+ n_i\,\nabla\cdot{\bf V}_i,\\[0.5ex]
{\bf 0} &= m_i\,n_i\,({\bf V}_i\cdot\nabla){\bf V}_i - e\,n_i\,({\bf E}+ {\bf V}_i\times{\bf B}) +\nabla\cdot \underline{\underline{P}}_{\,i}.\label{e6}
\end{align}
Here, ${\bf E}$ is the electric field-strength. 

 Equations~(\ref{e1})--(\ref{e4}) are the steady-state Maxwell
equations. Equation~(\ref{e4a}) is the steady-stage electron fluid continuity equation.
Equation~(\ref{e5}) is the steady-state, ideal equation of motion of the electron fluid, neglecting electron inertia, but incorporating an
anisotropic pressure tensor. Equation~(\ref{e6a}) is the ion fluid continuity equation.
 Finally, Eq.~(\ref{e6})
is the steady-state, ideal equation of motion of the ion fluid, including ion inertia, and incorporating an anisotropic pressure tensor.  Note that the ions are assumed to be singly charged. 
The fact that we are assuming 
 that $T_{\parallel\,s}$ and $T_{\perp\,s}$ are specified on a given magnetic field-line  negates the need for electron and
 ion energy conservation equations. 

\section{Normalization}
Let $L$ be a typical equilibrium variation length-scale (which is assumed to be similar to the physical dimensions of the mirror), $B_0$ a typical magnetic field-strength,  $n_c$ a typical number density, and $T_c$ a typical temperature.  The mean ion thermal velocity is $v_i=(T_c/m_i)^{1/2}$. The mean ion gyro-frequency is ${\mit\Omega}_{g\,i}=e\,B_0/m_i$. The ion gyro-radius is $\rho_i= v_i/{\mit\Omega}_{g\,i}$. 
Let $\nabla = L^{-1}\,\hat{\nabla}$, $\nabla_\parallel = L^{-1}\,\hat{\nabla}_\parallel$, $n_s=n_c\,\hat{n}_s$, 
$T_{\perp\,s} = T_c\,\hat{T}_{\perp\,s}$, $T_{\parallel\,s} = T_c\,\hat{T}_{\parallel\,s}$, 
${\bf V}_s= v_i\,\hat{\bf V}_s$, ${\bf B}= B_0\,\hat{\bf B}$, 
and ${\bf E} = B_0\,v_i\,\hat{\bf E}$. These orderings are consistent with an axisymmetric mirror that is rotating close to the ion thermal speed. 
 Equations~(\ref{e1})--(\ref{e6}) become
\begin{align}
\left(\frac{\lambda_D}{\rho_i}\right)^2 \hat{\rho}_i\,\hat{\nabla}\cdot\hat{\bf E}& = \hat{n}_i-\hat{n}_e,\label{e8}\\[0.5ex]
\hat{\nabla}\cdot\hat{\bf B} &= 0,\label{e9}\\[0.5ex]
\hat{\nabla}\times\hat{\bf E} &= {\bf 0}\label{e10},\\[0.5ex]
\hat{\rho}_i\,\hat{\nabla}\times\hat{\bf B} &= \beta\,(\hat{n}_i\,\hat{\bf V}_i-\hat{ n}_e\,\hat{\bf V}_e),\label{e11}
\end{align}
and 
\begin{align}\label{e12a}
0 &= \hat{\bf V}_e\cdot\hat{\nabla}\ln \hat{n}_e + \hat{\nabla}\cdot\hat{\bf V}_e,\\[0.5ex]
{\bf 0} &=\hat{\bf E} +\hat{\bf V}_e\times\hat{\bf B}\nonumber\\[0.5ex]
&\phantom{=}+\hat{\rho}_i\,\hat{T}_{\parallel\,e}\left[\hat{\nabla}_\parallel\ln \hat{n}_e+\gamma_e\,\hat{\nabla}_\parallel \ln \hat{B}
 +(1+\gamma_e)\,\hat{\nabla}_\perp\ln(\hat{n}_e\,\hat{T}_{\parallel\,e}) \right.\nonumber\\[0.5ex]
&\phantom{=}\left.
+\hat{\nabla}_\perp\gamma_e-\gamma_e\,\boldsymbol{\kappa}_\perp\right],\label{e12}
\end{align}
and
\begin{align} 
0 &= \hat{\bf V}_i\cdot\hat{\nabla}\ln\hat{n}_i + \hat{\nabla}\cdot\hat{\bf V}_i,\label{e13a}\\[0.5ex]
{\bf 0}&= -\hat{\bf E} - \hat{\bf V}_i\times\hat{\bf B}+\hat{\rho}_i\,(\hat{\bf V}_i\cdot\hat{\nabla})\hat{\bf V}_i 
\nonumber\\[0.5ex]
&\phantom{=}+\hat{\rho}_i\,\hat{T}_{\parallel\,i}\left[\hat{\nabla}_\parallel\ln \hat{n}_i+\gamma_i\,\hat{\nabla}_\parallel \ln \hat{B}
 +(1+\gamma_i)\,\hat{\nabla}_\perp\ln(\hat{n}_i\,\hat{T}_{\parallel\,i}) \right.
\nonumber\\[0.5ex]&\phantom{=}\left.+\hat{\nabla}_\perp\gamma_i-\gamma_i\,\boldsymbol{\kappa}_\perp\right], \label{e13}
\end{align}
where
\begin{align}
\lambda_D &= \left(\frac{\epsilon_0\,T_c}{n_c\,e^2}\right)^{1/2}, \\[0.5ex]
\beta &= \frac{\mu_0\,n_c\,T_c}{B_0^{\,2}}, 
\end{align}
and $\hat{\rho}_i=\rho_i/L$. Here, $\lambda_D$ is the Debye length, and $\beta$ is the ratio of the plasma thermal energy density to the magnetic energy density [\cite{Fitzpatrick2023}].

\section{Electric and magnetic fields}
Let $r$, $\theta$, $z$ be cylindrical coordinates. In an axisymmetric magnetic mirror, 
\begin{equation}
\frac{\partial}{\partial\theta} \equiv 0
\end{equation}
and $\hat{B}_\theta = 0$.
We can satisfy Eq.~(\ref{e9}) by writing [\cite{Hazeltine2026}]
\begin{equation}\label{eb}
\hat{\bf B} = \hat{\nabla}\alpha\times \hat{\nabla}\theta,
\end{equation}
where $\alpha(r,z)$ is a dimensionless magnetic field-line label. Thus [\cite{Hazeltine2026}],
\begin{equation}\label{e21}
\hat{\nabla}\times\hat{\bf B} = -(\hat{\mit\Delta}^\ast\alpha)\,\hat{\nabla}\theta = -\frac{\hat{\mit\Delta}^\ast\alpha}{\hat{r}}\,{\bf e}_\theta,
\end{equation}
where
\begin{equation}
\hat{\mit\Delta}^\ast \equiv \frac{\partial^2}{\partial\hat{r}^{\,2}}-\frac{1}{\hat{r}}\,\frac{\partial}{\partial\hat{r}}+ \frac{\partial^2}{\partial\hat{z}^{\,2}},
\end{equation}
and $\hat{r}=r/L$, $\hat{z}=z/L$. In an axisymmetric equilibrium, our previous assumptions imply that $\hat{T}_{\parallel\,s} =\hat{T}_{\parallel\,s}(\alpha)$,
$\hat{T}_{\perp\,s} =\hat{T}_{\perp\,s}(\alpha)$, and $\gamma_s=\gamma_s(\alpha)$. Note that we could, alternatively, have assumed
that  $\hat{T}_{\parallel\,s} =\hat{T}_{\parallel\,s}(\alpha,B)$,
$\hat{T}_{\perp\,s} =\hat{T}_{\perp\,s}(\alpha,B )$ and $\gamma_s=\gamma_s(\alpha,B )$. Although  this assumption would be more realistic, it would complicate the analysis, without
significantly modifying any of the main conclusions of the paper. 

We can satisfy Eq.~(\ref{e10}) by writing
\begin{equation}\label{e23}
\hat{\bf E} = -\hat{\nabla}\hat{\phi},
\end{equation}
where $\phi = v_i\,B_0\,L\,\hat{\phi}$ is the electrostatic potential. 

\section{Ordering}
Our fundamental ordering scheme  is
\begin{equation}
\lambda_D \ll \rho_i \ll L.
\end{equation}
In other words, the Debye length is much less than the ion gyro-radius, which, in turn, is much less than the machine dimensions. 
It follows that
\begin{align}
\frac{\lambda_D}{\rho_i}& \ll 1,\\[0.5ex]
\hat{\rho}_i &\ll 1.
\end{align}
Let $\beta$ be order unity. Thus, all quantities appearing in Eqs.~(\ref{e8})--(\ref{e13}) have leading-order contributions that are of order unity,
except for $\hat{\rho}_i$ and $\lambda_D/\rho_i$, which are both much less than unity. 

Up to first order in small quantities, Eq.~(\ref{e8}) gives the quasi-neutrality constraint 
\begin{equation}\label{e27}
\hat{n}_i= \hat{n}_e.
\end{equation}
Let
\begin{align}\label{e32}
\hat{\bf V}_i &= \hat{\bf V}_i^{(0)} + \hat{\rho}_i\,\hat{\bf V}_i^{(1)},\\[0.5ex]
\hat{\bf V}_e&= \hat{\bf V}_e^{(0)} + \hat{\rho}_i\,\hat{\bf V}_e^{(1)},\\[0.5ex]
\hat{\phi} &= \hat{\phi}^{(0)}+\hat{\rho}_i\,\hat{\phi}^{(1)},\label{e34}
\end{align}
where all superscripted quantities are order unity. 

\section{Lowest-order flow}
To lowest order in $\hat{\rho}_i$, Eqs.~(\ref{e12}), (\ref{e13}),  (\ref{e23}), and (\ref{e32})--(\ref{e34})  give
\begin{align}
{\bf 0}&= -\hat{\nabla}\hat{\phi}^{(0)}+\hat{\bf V}_e^{(0)}\times\hat{\bf B},\\[0.5ex]
{\bf 0}&= -\hat{\nabla}\hat{\phi}^{(0)}+\hat{\bf V}_i^{(0)}\times\hat{\bf B}.
\end{align}
Thus, we deduce that 
\begin{equation}
\hat{\bf B}\cdot\hat{\nabla}\hat{\phi}^{(0)} = 0,
\end{equation}
which implies that
\begin{equation}\label{e40g}
\hat{\phi}^{(0)} = {\mit\Phi}(\alpha).
\end{equation}
Hence, 
\begin{equation}
\hat{\bf V}_i^{(0)} = \hat{\bf V}_e^{(0)} = \frac{{\mit\Phi}'\,\hat{\nabla}\theta}{|\hat{\nabla}\theta|^2}= \hat{r}\,{\mit\Phi}'\,{\bf e}_\theta,
\end{equation}
where ${\mit\Phi}'\equiv d{\mit\Phi}/d\alpha$, use has been made of the fact that $|\hat{\nabla}\theta|=1/\hat{r}$, and the zeroth-order parallel (to the magnetic field) components of $\hat{\bf V}_i$ and $\hat{\bf  V}_e$ are assumed to be zero. 
It follows that
\begin{equation}\label{e42g}
\hat{\mit\Omega}_\theta\equiv\hat{\bf V}_i^{(0)}\cdot{\bf e}_\theta/\hat{r} = \hat{\bf V}_e^{(0)}\cdot{\bf e}_\theta/\hat{r} ={\mit\Phi}',
\end{equation}
where ${\mit\Omega}_\theta = \hat{\mit\Omega}_\theta \,v_i/L$ is the lowest-order plasma angular velocity. 
Thus, to lowest order, the electrostatic potential is constant along magnetic field-lines, and 
 both charge species rotate with a common  angular velocity that  is also constant along magnetic field-lines [\cite{Hinton1985}]. In other words, 
 $\hat{\mit\Omega}_\theta=\hat{\mit\Omega}_\theta(\alpha)=d{\mit\Phi}/d\alpha$,
which implies that the flow is {\em iso-rotational}: i.e., $\hat{\bf B}\cdot\hat{\nabla}{\mit\Omega}_\theta=0$ [\cite{Ferraro1937,Lehnert1971,Hinton1985}].
It also follows that
\begin{equation}\label{e38}
(\hat{\bf V}_i^{(0)}\cdot\hat{\nabla})\hat{\bf V}_i^{(0)} = - \hat{r}\,\hat{\mit\Omega}_\theta^{\,2}\,{\bf e}_r.
\end{equation}
Note that 
\begin{equation}
\hat{\bf V}_e^{(0)}\cdot\hat{\nabla}\ln\hat{n}_e=\hat{\nabla}\cdot\hat{\bf V}_e^{(0)}= \hat{\bf V}_i^{(0)}\cdot\hat{\nabla}\ln\hat{n}_i=\hat{\nabla}\cdot\hat{\bf V}_i^{(0)}=0,
\end{equation}
which ensures that the zeroth-order components of the electron and ion continuity equations, (\ref{e12a}) and (\ref{e13a}), respectively, are
automatically satisfied.

\section{Parallel force balance}
Now, from Eq.~(\ref{eb}), 
\begin{equation}
\hat{\bf B}\cdot\hat{\nabla} = {\cal J}^{-1}\,\frac{\partial}{\partial s},
\end{equation}
where $s$ is an element of normalized arc-length along a magnetic field-line, and ${\cal J} = (\hat{\nabla}\alpha\times\hat{\nabla}\theta\cdot\hat{\nabla}{s})^{-1}$ is the Jacobian of the $\alpha$, $\theta$, $s$ coordinate system. Here, $\partial/\partial s$ is taken at constant $\alpha$ and $\theta$ 

If we take the dot product of the first-order (in $\hat{\rho}_i$) version of Eq.~(\ref{e12}) with $\hat{\bf B}$, then we get
\begin{equation}\label{e46y}
0=\frac{\partial}{\partial s}\!\left[-\hat{\phi}^{(1)}+\hat{T}_{\parallel\,e}\,(\ln\hat{n}_e+\xi_e)\right],
\end{equation}
where
\begin{equation}
\xi_s = \gamma_s\,\ln \hat{B}.
\end{equation}
It follows that [\cite{Catto1987}]
\begin{equation}\label{e40}
\hat{T}_{\parallel\,e}\,\left(\ln\hat{n}_e+\xi_e\right) = \hat{\phi}^{(1)}+ f(\alpha),
\end{equation}
where $f(\alpha)$ is arbitrary. 
Likewise, if we take the dot product of the first-order version of Eq.~(\ref{e13}) with $\hat{\bf B}$, then, making use of Eq.~(\ref{e38}), we get
\begin{equation}
0=-\frac{\hat{r}\,\hat{\mit\Omega}_\theta^{\,2}}{|\hat{\nabla}\hat{r}|}\,\frac{\partial\hat{r}}{\partial s} +\frac{\partial}{\partial s}\!\left[\hat{\phi}^{(1)}
+\hat{T}_{\parallel\,i}\,(\ln \hat{n}_i +\xi_i)\right],
\end{equation}
or
\begin{equation}
0=\frac{\partial}{\partial s}\!\left[-\frac{1}{2}\,\hat{r}^{\,2}\,\hat{\mit\Omega}_\theta^{\,2} +\hat{\phi}^{(1)}+ \hat{T}_{\parallel\,i}\,(\ln\hat{n}_i+\xi_i)\right],
\end{equation}
where use has been made of the facts that $\hat{\mit\Omega}_\theta=\hat{\mit\Omega}_\theta(\alpha)$ and $|\hat{\nabla}\hat{r}|=1$. 
It follows that [\cite{Catto1987}]
\begin{equation}\label{e42}
\hat{T}_{\parallel\,i}\,(\ln\hat{n}_i+\xi_i) =  \frac{1}{2}\,\hat{r}^{\,2}\,\hat{\mit\Omega}_\theta^{\,2}-\hat{\phi}^{(1)}+ g(\alpha),
\end{equation}
where $g(\alpha)$ is arbitrary. 

Let us define the field-line average operator [\cite{Hazeltine2026}]
\begin{equation}
\langle F\rangle(\alpha) \equiv \left.\int^s F(\alpha,s)\,ds\right/\int^s ds, 
\end{equation}
where the integral is taken along the whole magnetic field-line whose label is $\alpha$ (i.e., from end-plate to end-plate), and let
\begin{equation}
\tilde{F}\equiv  F-\langle F\rangle.
\end{equation}
Without loss of generality, we can set
\begin{equation}\label{e46}
 \langle \hat{\phi}^{(1)}\rangle=0,
\end{equation}
which allows us to determine the two functions $f(\alpha)$ and $g(\alpha)$. 

The quasi-neutrality constraint, (\ref{e27}), can be combined with Eqs.~(\ref{e40}), (\ref{e42}), and (\ref{e46}) to give [\cite{Hinton1985,Schwatz2024}]
 \begin{equation}\label{e56}
 \hat{\phi}^{(1)}(\alpha,s) = \hat{T}_{\parallel\,e}\,(\hat{T}_{\parallel\,e}+ \hat{T}_{\parallel\,i})^{\,-1}\left[\frac{1}{2}\,\widetilde{\hat{r}^{\,2}}\,\hat{\mit\Omega}_\theta^{\,2}+\hat{T}_{\parallel\,i}\,(\tilde{\xi}_e-\tilde{\xi}_i)\right],
 \end{equation}
 where [\cite{Hinton1985}]
 \begin{align}\label{e57}
 \hat{n}_i(\alpha,s) &= \hat{n}_e(\alpha,s)= \hat{n}_0(\alpha)\exp
 \left[(\hat{T}_{\parallel\,e}+ \hat{T}_{\parallel\,i})^{\,-1}\left(\frac{1}{2}\,\widetilde{\hat{r}^{\,2}}\,\hat{\mit\Omega}_\theta^{\,2}-\hat{T}_{\parallel\,e}\,\tilde{\xi}_e-\hat{T}_{\parallel\,i}\,\tilde{\xi}_i\right)\right],\\[0.5ex]
 f(\alpha) &=\hat{T}_{\parallel\,e}\left(\ln\hat{n}_0+\langle\xi_e\rangle\right),\\[0.5ex]
 g(\alpha)&= \hat{T}_{\parallel\,i}\,(\ln\hat{n}_0 + \langle\xi_i\rangle)- \frac{1}{2}\,\langle\hat{r}^{\,2}\rangle\,\hat{\mit\Omega}_\theta^{\,2}. 
 \end{align}
 It follows that [\cite{Hinton1985,Catto1987}]
 \begin{align}
 \hat{n}_e(\alpha,s) &= \hat{n}_0(\alpha)\,\exp\left(\hat{T}_{\parallel\,e}^{\,-1}\,\hat{\phi}^{(1)} - \tilde{\xi}_e\right),\label{e61g}\\[0.5ex]
 \hat{n}_i(\alpha,s) &= \hat{n}_0(\alpha)\,\exp\left[\hat{T}_{\parallel\,i}^{\,-1}\left(\frac{1}{2}\,\widetilde{\hat{r}^{\,2}}\,\hat{\mit\Omega}_\theta^{\,2}-\tilde{\phi}^{(1)}\right) -\tilde{\xi}_i\right].\label{e62g}
 \end{align}
 According to Eq.~(\ref{e56}), a nonuniform parallel electric field develops on a given magnetic field-line. This field ensures that the
 previous two Boltzmann-like expressions yield $\hat{n}_e(\alpha,s)=\hat{n}_i(\alpha,s)$,  as is required by the quasi-neutrality constraint (\ref{e27}). 
 
Equation~(\ref{e57}) can be rewritten in the form
 \begin{align}\label{e57a}
& \hat{n}_i(\alpha,s) = \hat{n}_e(\alpha,s)\\[0.5ex] &=  \hat{n}_0(\alpha)\exp\left\{-(\hat{T}_{\parallel\,e}+\hat{T}_{\parallel\,i})^{\,-1}\,\left
[({\mit\Delta}\hat{T}_{e}+{\mit\Delta}\hat{T}_i)\left(\ln\hat{B}-\langle\ln\hat{B}\rangle\right)+
\frac{1}{2}\,\hat{\mit\Omega}_\theta^{\,2}\left(\langle\hat{r}^{\,2}\rangle-\hat{r}^{\,2}\right)
 \right]\right\},\nonumber
 \end{align}
 where ${\mit\Delta}\hat{T}_s =\hat{T}_{\perp\,s} - \hat{T}_{\parallel\,s}$. 
 By examining this equation, we can appreciate the two possible types of plasma confinement in an axisymmetric mirror device. Conventional particle trapping (i.e., the terms involving ${\mit\Delta}\hat{T}_s$) reduces the relative
 particle density on those parts of a magnetic field-line where the magnetic field-strength is greater than average [\cite{Endrizii2023}]. Centrifugal confinement, on the other hand, (i.e.,  the term involving $\hat{\mit\Omega}_\theta^{\,2}$) reduces the particle density on those parts of a magnetic field-line that are closer than average to the magnetic
 axis [\cite{Schwatz2024}]. 
 
 Note, finally, that in the absence of plasma rotation, Eq.~(\ref{e57a}) can be written
 \begin{equation}
  \hat{n}_i(\alpha,B) = \hat{n}_e(\alpha,B)= \hat{n}_0(\alpha)\left(\frac{\bar{B}}{B}\right)^\zeta,
  \end{equation}
  where
  \begin{align}
  \zeta(\alpha) &= \frac{{\mit\Delta}\hat{T}_{e}+{\mit\Delta}\hat{T}_i}{\hat{T}_{\parallel\,e}+\hat{T}_{\parallel\,i}},\\[0.5ex]
  \bar{B}(\alpha) &= \exp(\langle \ln B\rangle).
  \end{align}
  The idea that, in a non-rotating mirror plasma,  $\hat{n}_s=\hat{n}_s(\alpha,B)$, which also implies that $p_{\parallel\,s}= p_{\parallel\,s}(\alpha,B)$ and 
   $p_{\perp\,s}= p_{\perp\,s}(\alpha,B)$, and, moreover,  that the functional relation between the parallel and perpendicular pressures and the magnetic
   field-strength is a power law, is quite an old one [\cite{Taylor1963,Grad1967,Lindvall2025}]. 
   
\section{First-order flows}
 The first-order version  of Eq.~(\ref{e12}) yields
 \begin{align}
 \hat{\bf V}_e^{(1)}\times \hat{\bf B} &=\boldsymbol{\chi}_{\perp\,e},
 \end{align}
 where 
 \begin{align}
\boldsymbol{\chi}_{\perp\,e}&=-\hat{T}_{\parallel\,e}\left\{(1+\gamma_e)\,\hat{\nabla}_\perp\!\left[\ln(\hat{n}_0\,\hat{T}_{\parallel\,e})-\tilde{\xi}_e\right]
+\gamma_e\,\hat{\nabla}_\perp\!\left(\hat{T}_{\parallel\,e}^{\,-1}\,\hat{\phi}^{(1)}\right)
\right.\nonumber\\[0.5ex]
&\phantom{=}\left.+ \hat{\nabla}_\perp\gamma_e -\gamma_e\,\boldsymbol{\kappa}_\perp\right\},
 \end{align}
 and use has been made of Eqs.~(\ref{e56}), (\ref{e57}), and (\ref{e61g}). 
  Thus, we obtain
 \begin{equation}
 \hat{\bf V}_e^{(1)}= \frac{\hat{\bf B}\times\boldsymbol{\chi}_{\perp\,e}}{\hat{B}^{\,2}}+ \hat{V}_{\parallel\,e}^{(1)}\,{\bf b},
 \end{equation}
  which reduces to
 \begin{equation}
  \hat{\bf V}_e^{(1)}= \hat{r}\,\hat{\mit\Omega}_{\theta\,e}^{(1)}\,{\bf e}_\theta + \hat{V}_{\parallel\,e}^{(1)}\,{\bf b},
\end{equation}
where 
\begin{equation}
\hat{\mit\Omega}_{\theta\,e}^{(1)}(\alpha,s) = \frac{\hat{\nabla}\alpha\cdot\boldsymbol{\chi}_{\perp\,e}}{|\hat{\nabla} \alpha|^2}.
\end{equation}

The first-order version of Eq.~(\ref{e12a}) gives
\begin{equation}
0 = \hat{\bf V}_e^{(1)}\cdot\hat{\nabla}\ln\,\hat{n}_e+\hat{\nabla}\cdot\hat{\bf V}_e^{(1)},
\end{equation}
 which reduces to
 \begin{equation}\label{e62x}
 \hat{\bf B}\cdot\hat{\nabla}\left[\ln\left(\frac{\hat{V}_{\parallel\,e}^{(1)}\,\hat{n}_e}{\hat{B}}\right)\right]= 0.
 \end{equation}
 Making use of Eq.~(\ref{e57}), we deduce that
 \begin{equation}\label{e63x}
 \frac{\hat{V}_{\parallel\,e}^{(1)}}{\hat{B}} = \hat{V}_{\parallel\,e\,0}^{(1)}(\alpha)\exp\left[-
 (\hat{T}_{\parallel\,e}+ \hat{T}_{\parallel\,i})^{\,-1}\left(\frac{1}{2}\,\widetilde{\hat{r}^{\,2}}\,\hat{\mit\Omega}_\theta^{\,2}-\hat{T}_{\parallel\,e}\,\tilde{\xi}_e-\hat{T}_{\parallel\,i}\,\tilde{\xi}_i\right)
\right],
 \end{equation}
 where $\hat{V}_{\parallel\,e\,0}^{(1)}(\alpha)$ is arbitrary. 
 Note that the parallel electron fluid velocity, $\hat{V}_{\parallel\,e}^{(1)}$, cannot change sign on a given magnetic field-line. In this paper, we restrict attention to equilibria without imposed end-to-end particle fluxes. Thus, we set
 $\hat{V}_{\parallel\,e\,0}^{(1)}(\alpha)=0$ on all field-lines, which implies that $\hat{V}_{\parallel\,e}^{(1)}=0$ on all field-lines. 

The first-order version of Eq.~(\ref{e13}) gives
\begin{align}
\hat{\bf V}_i^{(1)}\times \hat{\bf B} &=\boldsymbol{\chi}_{\perp\,i},
\end{align}
where
\begin{align}
\boldsymbol{\chi}_{\perp\,i}& = 
\hat{T}_{\parallel\,i}\left\{(1+\gamma_i)\,\hat{\nabla}_\perp\!\left[\ln(\hat{n}_0\,\hat{T}_{\parallel\,i})-\tilde{\xi}_i\right]
-\gamma_i\,\hat{\nabla}_\perp\!\left(\hat{T}_{\parallel\,i}^{\,-1}\,\hat{\phi}^{(1)}\right)
\right.\nonumber\\[0.5ex]
&\phantom{=}\left.+ \hat{\nabla}_\perp \gamma_i -\gamma_i\,\boldsymbol{\kappa}_\perp\right\}
\\[0.5ex]
&\phantom{=}
-\frac{1}{2}\,\hat{\mit\Omega}_{\theta}^{\,2}\,\hat{\nabla}_\perp\langle\hat{r}^{\,2}\rangle + \frac{1}{2}\,\widetilde{\hat{r}^{\,2}}\,\hat{\nabla}_\perp
\hat{\mit\Omega}_\theta^{\,2} + \gamma_i\,\hat{\nabla}_\perp\!\left(\frac{1}{2}\,\widetilde{\hat{r}^{\,2}}\,\hat{\mit\Omega}_\theta^{\,2}\right)
-\gamma_i\,\frac{1}{2}\,\widetilde{\hat{r}^{\,2}}\,\hat{\mit\Omega}_\theta^{\,2}\,\hat{\nabla}_\perp\ln\hat{T}_{\parallel\,i},\nonumber
\end{align}
and
 use has been made of Eqs.~(\ref{e38}), (\ref{e56}), (\ref{e57}), and (\ref{e62g}). 
 Thus, we obtain
 \begin{equation}
 \hat{\bf V}_i^{(1)}= \frac{\hat{\bf B}\times\boldsymbol{\chi}_{\perp\,i}}{\hat{B}^{\,2}}+ \hat{V}_{\parallel\,i}^{(1)}\,{\bf b},
 \end{equation}
  which reduces to
 \begin{equation}
  \hat{\bf V}_i^{(1)}= \hat{r}\,\hat{\mit\Omega}_{\theta\,i}^{(1)}\,{\bf e}_\theta + \hat{V}_{\parallel\,i}^{(1)}\,{\bf b},
\end{equation}
where 
\begin{equation}
\hat{\mit\Omega}_{\theta\,i}^{(1)}(\alpha,s) = \frac{\hat{\nabla}\alpha\cdot\boldsymbol{\chi}_{\perp\,i}}{|\hat{\nabla} \alpha|^2}.
\end{equation}
 
 The first-order version of Eq.~(\ref{e13a}) gives
\begin{equation}
0 = \hat{\bf V}_i^{(1)}\cdot\hat{\nabla}\ln\,\hat{n}_i+\hat{\nabla}\cdot\hat{\bf V}_i^{(1)},
\end{equation}
 which reduces to
 \begin{equation}
 \hat{\bf B}\cdot\hat{\nabla}\left[\ln\left(\frac{\hat{V}_{\parallel\,i}^{(1)}\,\hat{n}_i}{\hat{B}}\right)\right]= 0.
 \end{equation}
 By analogy with Eqs.~(\ref{e62x}) and (\ref{e63x}), we deduce that
 \begin{equation}
 \frac{\hat{V}_{\parallel\,i}^{(1)}}{\hat{B}}=\hat{V}_{\parallel\,i\,0}^{(1)}(\alpha)\exp\left[
 -
 (\hat{T}_{\parallel\,e}+ \hat{T}_{\parallel\,i})^{-1}\left(\frac{1}{2}\,\widetilde{\hat{r}^{\,2}}\,\hat{\mit\Omega}_\theta^{\,2}-\hat{T}_{\parallel\,e}\,\tilde{\xi}_e-\hat{T}_{\parallel\,i}\,\tilde{\xi}_i\right)
\right].
 \end{equation}
 As before, because we are restricting our attention to equilibria without imposed end-to-end particle fluxes,  we set
 $\hat{V}_{\parallel\,i}^{(1)}=0$ on all field-lines. 

It follows from the previous analysis that the ion and electron fluid velocities can be written
\begin{align}\label{e67}
\hat{\bf V}_e  &= \hat{r}\,\hat{\mit\Omega}_e\,{\bf e}_\theta.\\[0.5ex]
\hat{\bf V}_i  &= \hat{r}\,\hat{\mit\Omega}_i\,{\bf e}_\theta,
\end{align}
In other words, the ions and electron fluids both circulate about the $\theta$-axis with (normalized) angular velocities, $\hat{\mit\Omega}_e$ and
$\hat{\mit\Omega}_i$, respectively, 
where
\begin{align}\label{e69}
\hat{\mit\Omega}_e(\alpha,s) &= \hat{\mit\Omega}_\theta(\alpha) +\hat{\rho}_i\,\hat{\mit\Omega}_{\theta\,e}^{(1)}(\alpha,s),\\[0.5ex]
\hat{\mit\Omega}_i(\alpha,s)&=\hat{\mit\Omega}_\theta(\alpha)+ \hat{\rho}_i\,\hat{\mit\Omega}_{\theta\,i}^{(1)}(\alpha,s).\label{e70}
\end{align}

\section{Grad-Shafranov equation}
 Equation~(\ref{e11}) yields the Grad-Shafranov equation
 \begin{equation}\label{e55}
\hat{\mit\Delta}^\ast\alpha = \beta\,\hat{r}^{\,2}\,\hat{n}_0\,\exp\left(\hat{T}_{\parallel\,e}^{\,-1}\,\hat{\phi}^{(1)} - \tilde{\xi}_e\right)\left[{\mit\Omega}_{\theta\,e}^{(1)} - {\mit\Omega}_{\theta\,i}^{(1)}\right], 
 \end{equation}
 where use has been made of Eqs.~(\ref{e21}),  (\ref{e61g}), and (\ref{e67})--(\ref{e70}). The previous equation describes how the $\theta$-directed plasma
 current that flows in the plasma modifies the structure of the magnetic field. It is clear from our previous expressions for ${\mit\Omega}_{\theta\,e}^{(1)}$  and ${\mit\Omega}_{\theta\,i}^{(1)}$ that the
 current has a diamagnetic component driven by perpendicular pressure gradients, a component driven by
 temperature anisotropy, a component driven by the lowest-order plasma rotation, and a component driven by magnetic field-line curvature. 
  
\section{Summary and discussion}
We have investigated the equilibrium of a rotating magnetic mirror using ideal two-fluid theory. Our main results are as follows. 
  Equation~(\ref{e40g}) states that the lowest-order electrostatic potential is constant on a particular magnetic field-line. We expect this potential
 to be determined by the spatial boundary conditions. 
 Equation~(\ref{e56})  states how the first-order electrostatic potential, which is
 needed to ensure quasi-neutrality, varies along a particular field-line due to centrifugal and temperature anisotropy effects.  Equation~(\ref{e57a}) specifies how the common number density of the two charge species varies along magnetic field-lines, due to centrifugal and temperature anisotropy effects. 
 According to Eqs.~(\ref{e69}) and (\ref{e70}), the fact that the  lowest-order electrostatic potential is constant on a given magnetic field-line  implies that both charge species  rotate at a common angular velocity that is  also constant on magnetic field-lines. To higher order, the two charge species have slightly different angular velocities, due to diamagnetic, centrifugal,
 temperature anisotropy, and magnetic field-line curvature effects. Moreover, these higher-order corrections are not necessarily constant on magnetic field-lines. Finally, Eq.~(\ref{e55}) can be solved to determine the structure of the equilibrium magnetic field, once the vacuum magnetic field, as well as the 
 arbitrary functions $\hat{n}_0(\alpha)$, $\hat{\mit\Omega}_\theta(\alpha)$, $\hat{T}_{\parallel\,e}(\alpha)$, $\hat{T}_{\perp\,e}(\alpha)$,
 $\hat{T}_{\parallel\,i}(\alpha)$, and $\hat{T}_{\perp\,i}(\alpha)$,  are specified. 
  
The ordering assumptions which underlie our analysis apply as long as the ion gyro-radius is much less than the size of the machine  (which is assumed to
be the typical length-scale on which equilibrium quantities vary), and 
as long as first-order terms in our expressions for the electrostatic potential and the angular velocities of the
two charge species remain much smaller than zeroth-order terms. This is the case provided
\begin{equation}
\hat{\mit\Omega}_\theta \ll \frac{1}{\hat{\rho}_i},
\end{equation}
or
\begin{equation}\label{e59}
{\mit \Omega}_\theta \ll {\mit\Omega}_{g\,i}.
\end{equation}
Obviously, it would not be possible to confine ions in the mirror if the ion gyro-radius were greater than, or of order, the machine size.
Likewise, it seem completely impractical to spin the plasma at an angular velocity, ${\mit\Omega}_\theta$,  that is greater than, or of order, the
ion gyro-frequency,  ${\mit\Omega}_{g\,i}$.   (For one thing, spinning a plasma this rapidly would lead to the complete breakdown of the guiding-center theory which is the whole basis for
ion confinement in a mirror.) Thus, our orderings apply to any practical mirror machine with sonic [or even supersonic, provided that the criterion (\ref{e59}) is not violated] rotation. 
  
The main question addressed in this paper is whether  the standard result, (ultimately) due to Ferraro, 
that the plasma angular
velocity is approximately constant along individual equilibrium magnetic field-lines in a rotating axisymmetric magnetic mirror machine, 
continues to hold when the rotation becomes sonic or supersonic.
Equation~(\ref{e40g}) shows that, provided our ordering assumptions are satisfied,  the lowest-order electrostatic potential in a rotating magnetic mirror machine is constant on magnetic field-lines,
which implies, from Eq.~(\ref{e42g}), that the lowest-order plasma angular velocity is also constant on magnetic field-lines, in accordance with [\cite{Ferraro1937}]. Our analysis can be reconciled with that of  \cite{Hazeltine2026}
by noting  that the solution in the latter paper that is termed ``subsonic", which  is similar to the solution described in this paper, actually applies to all plasma angular velocities that are significantly less than the ion gyro-frequency
(which implies that it also applies to moderately supersonic rotation), whereas the solution that is termed ``sonic" only applies to impractically high rotation levels such
that the plasma angular velocity is similar to the ion gyro-frequency. 

Of course, an axisymmetric mirror confinement device differs fundamentally from a toroidal confinement system because all of the
magnetic field-lines that confine the plasma eventually encounter the two conducting  end-plates that bookend the device. We expect very thin plasma sheaths
(whose thicknesses are of order the Debye length) to form on the inward facing surfaces of the end-plates. However, we would also expect the vast difference in scale between the
thicknesses of the sheaths and the dimensions of the main plasma to allow the sheath solutions to effectively decouple from the main equilibrium solution. In other words, the
sheath solutions should only depend on the local properties of the main equilibrium solution at the two end-plates. As is well known, an electrostatic potential difference develops across a sheath. Moreover, this potential difference,
which is easily calculated,  depends, amongst other things, on the electron temperature and the ion sound speed [\cite{Reimann1991}].
Thus, in principle, by electrically biasing insulated segments of the two end-plates, it should be possible to bias individual magnetic field-lines within the main plasma [\cite{Endrizii2023}]. 
In other words, it  should be possible to adjust the lowest-order electrostatic potential, ${\mit\Phi}(\alpha)$, in the main plasma,
and, thereby, to vary the lowest-order plasma angular velocity, ${\mit\Omega}_\theta(\alpha)\propto d{\mit\Phi}/d\alpha$ [\cite{Endrizii2023}].


\section*{Funding}
This research was supported by the U.S.\ Department of Energy, Office of Science, Office of Fusion Energy Sciences,  under  contract DE-FG02-04ER54742. 


\end{document}